\documentclass[12pt]{article}

\usepackage{amssymb}
\usepackage{epsf}

\def\mprp{\mbox{\tiny $\bot$}}
\def\mprl{\mbox{\tiny $\|$}}

\title{
\begin{flushright}
{\normalsize Yaroslavl State University\\
             Preprint YARU-HE-02/02\\
             hep-ph/0204168} \\[5mm]
\end{flushright}
Electron mass operator \\ in a strong magnetic field}

\author{{A.V.~Kuznetsov, N.V.~Mikheev and M.V.~Osipov}\\ [7mm] 
{\small\it Division of Theoretical Physics, Department of Physics,}\\
{\small\it Yaroslavl State University, Sovietskaya 14,}\\
{\small\it 150000 Yaroslavl, Russian Federation}}

\date{}

\begin{document}

\maketitle

\begin{abstract}

The electron mass operator in a strong magnetic field is calculated 
by summation of the leading log contributions in all orders of 
the perturbation theory. An influence of the strong field 
on the virtual photon polarization operator is taken into account.
The contribution of higher Landau levels of virtual electrons, along 
with the ground Landau level, is shown to be essential in the leading 
log approximation.
\end{abstract}

\medskip


\medskip


\vfill

\begin{center}
{\sl 
Submitted to Modern Physics Letters A}
\end{center}

\newpage

Investigations of the asymptotic properties of diagrams and operators 
of the quantum electrodynamics in strong magnetic fields
$B \gg B_e \;(B_e = m_e^2/e \simeq 4.41 \cdot 10^{13}\;$ G~\footnote{
We use natural units in which $c = \hbar = 1$ and the pseudo-Euclidean 
metric with a $(+ - - -)$ signature. Hereafter $e$ is the elementary charge.}) 
is of conceptual interest both from the standpoint of the searches 
of the borders of the perturbation theory applicability, and 
in view of possible astrophysical applications.
The researches of this type are being performed by many authors during 
a rather long time. For example, a history of calculations of 
the electron mass operator in a strong magnetic field lasts more than 
30 years already. However, as we show in this letter, the problem has not 
been resolved yet.

The electron mass operator in a strong magnetic field with the one-loop 
contribution taken into account obtained by Jancovici~\cite{Jancovici:1969} 
in the leading log approximation can be written as
\begin{eqnarray}
M^{(1)} = m\, \left( 1 + \frac{\alpha}{4 \pi} \, \ln^2 \frac{e B}{m^2} 
\right).
\label{eq:Janc}
\end{eqnarray}
The origin of the double logarithm asymptotics can be easily illustrated 
by using of the electron propagator in a strong magnetic 
field~\cite{Loskutov:1976} for presenting the one-loop contribution 
in the form of the following integral
\begin{eqnarray}
\Delta M^{(1)} = - \, \frac{i \alpha}{2 \pi^3} \, m \, \left .
\int \, d^2 k_{\mprp} \, e^{- k_{\mprp}^2/2 e B} \; 
\int \, \frac{d^2 k_{\mprl}}{[(k - p)_{\mprl}^2 - m^2]
(k_{\mprl}^2 - k_{\mprp}^2)} \;\right\vert_{p_{\mprl}^2 = m^2},
\label{eq:M1_1}
\end{eqnarray}
where the notations are used:
$k_{\mprp}^2 = k_1^2 + k_2^2, \quad k_{\mprl}^2 = k_0^2 - k_3^2$,
(the magnetic field is directed along the 3d axis). 
The Wick rotation in the complex plane of $k_0\;$,
$k_0 = i {\tilde k_0}$ yields 
$k_{\mprl}^2 = - {\tilde k_0}^2 - k_3^2 \equiv - {\tilde k}_{\mprl}^2$.
With the main contribution into the integral coming from the region
$k_{\mprp}^2 \gg {\tilde k}_{\mprl}^2 \gg m^2$, 
one obtains
\begin{eqnarray}
\Delta M^{(1)} \simeq \frac{\alpha}{2 \pi} \, m \, 
\int\limits_{m^2}^\infty \, \frac{d k_{\mprp}^2}{k_{\mprp}^2} \, 
e^{- k_{\mprp}^2/2 e B} \; 
\int\limits_{m^2}^{k_{\mprp}^2} \, \frac{d {\tilde k}_{\mprl}^2}
{{\tilde k}_{\mprl}^2}.
\label{eq:M1_2}
\end{eqnarray}
It is seen that the result contains the logarithm squared, and coinsides 
with the second term of eq.~(\ref{eq:Janc}).

\begin{figure}[htb]

\epsfxsize=.8\textwidth
\epsfysize=.4\textwidth
\epsffile[0 0 488 128]{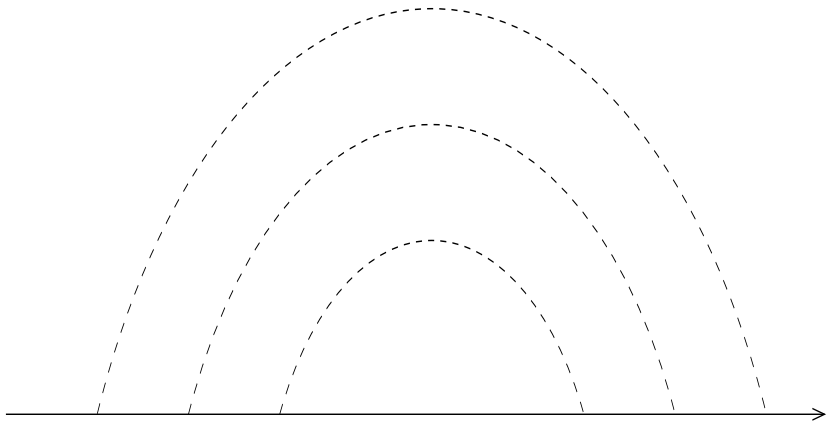}

\vspace{10mm}
\caption{The rainbow graph.}
\label{fig:rainbow}

\end{figure}

Later on, in the papers by Loskutov and 
Skobelev~\cite{Loskutov:1979,Loskutov:1981} the attempts were performed to 
calculate the two-loop contribution and to summarize all the many-loop 
contributions in the same leading log approximation.  
A rather effective technique of ``two-dimensional electrodynamics'' 
in a strong magnetic field, developed by the authors was used
and it was shown that the main contribution into the electron mass operator 
arises from the rainbow Feynman diagram depicted in
Fig.~\ref{fig:rainbow}.
The result of summation over all orders of the perturbation theory 
was obtained in Ref.~\cite{Loskutov:1981} in the exponential form:
\begin{eqnarray}
M_{LS} = m\, \exp \left(\frac{\alpha}{4 \pi} \, \ln^2 \frac{e B}{m^2} \right).
\label{eq:LS_81}
\end{eqnarray}
However, in the recent paper by Gusynin and Smilga~\cite{Gusynin:1999} 
the summation over all orders of the perturbation theory was remade 
giving another result:
\begin{eqnarray}
M_{GS} = m \left/ \cos \left(\sqrt{\frac{\alpha}{2 \pi}} \, 
\ln \frac{e B}{m^2} \right) \right ..
\label{eq:GS_99}
\end{eqnarray}
Obviously, the double logarithmic terms are also collected here because
$x^2$ is the parameters of the $\cos x$ expansion.
The first terms of expansion over $\alpha$ of both equations, (\ref{eq:LS_81}) 
and~(\ref{eq:GS_99}), coinside and reproduce the one-loop result
(\ref{eq:Janc}) of the paper~\cite{Jancovici:1969}.
The difference manifests itself at the two-loop level where the 
corresponding contributions from Eqs. (\ref{eq:LS_81}) and (\ref{eq:GS_99})
have the form
\begin{eqnarray}
\Delta M^{(2)}_{LS} = m \, \frac{1}{32} \, \frac{\alpha^2}{\pi^2} \, 
\ln^4 \frac{e B}{m^2},
\label{eq:DM_LS}\\
\Delta M^{(2)}_{GS} = m \, \frac{5}{96} \, \frac{\alpha^2}{\pi^2} \, 
\ln^4 \frac{e B}{m^2}.
\label{eq:DM_GS}
\end{eqnarray}
From the computation standpoint, in the frame of the approximation 
used by the authors, we confirm the result (\ref{eq:DM_GS}). 
It is intriguing to compare the Eqs. (\ref{eq:DM_LS}) and (\ref{eq:DM_GS})
with the earlier result for the two-loop contribution calculated 
in Ref.~\cite{Loskutov:1979}, see Eq. (5) of that paper. 
The numerical coefficient presented there, 5/48, was erroneous, to be 
replaced to 1/32, as it was pointed out in the next 
paper~\cite{Loskutov:1981}. In fact, the true coefficient appeared 
to be 5/96.

However, both equations ~(\ref{eq:Janc}) and~(\ref{eq:DM_GS}) had 
a restricted area of application because they were obtained without 
taking account of a crucial influence of the strong magnetic field 
on the virtual photon polarization operator.
As was shown for the first time in Ref.~\cite{Loskutov:1983}, 
due to the photon polarization in the field, the double logarithm asymptotics 
is only valid at $B \lesssim B_e/\alpha$. For the larger field values, 
a photon acquires the effective mass $m_\gamma^2 = (2\alpha/\pi) e B$.
Turning back to Eq.~(\ref{eq:M1_1}), we should replace the factor
$k^2 = k_{\mprl}^2 - k_{\mprp}^2$ in the denominator arising from the 
massless photon propagator, to the expression $k^2 - m_\gamma^2$. 
It changes the result essentially, because in one of the two 
logarithms the electron mass is replaced to the photon mass, 
$$\ln(e B/m_\gamma^2) \sim \ln(1/\alpha),$$
so, the dependence of the electron mass operator on the field becomes 
not double but single logarithmic:
\begin{eqnarray}
\Delta M^{(1)} = m \, \frac{\alpha}{2 \pi} \, 
\left(\ln \frac{\pi}{\alpha} - \gamma_E \right) \,
\ln \frac{e B}{m^2}, 
\label{eq:DM_1}
\end{eqnarray}
where $\gamma_E = 0.577 \dots$ is the Euler constant. 

However, the influence of the strong magnetic field on the photon 
polarization properties should be considered in more details.
As was shown by Shabad, see, e.g.~\cite{Shabad:1988} where the 
earlier references can be found, the photon propagator in a magnetic 
field can be presented in the form
\begin{eqnarray}
{\cal D}_{\mu \nu} (k) = - i \sum\limits_{\lambda=1}^3 \;
\frac{1}{k^2 - {\cal P}^{(\lambda)}} \;
\frac{b_\mu^{(\lambda)} b_\nu^{(\lambda)}}
{\left(b^{(\lambda)}\right)^2} ,
\label{eq:D_munu}
\end{eqnarray}
where $b_\mu^{(\lambda)}$ are the eigenvectors and ${\cal P}^{(\lambda)}$ 
are the eigenvalues of the photon polarization operator, 
$$b_{\alpha}^{(1)}  =  ( q \varphi )_{\alpha}, \quad
b_{\alpha}^{(2)}  =  ( q {\tilde \varphi} )_{\alpha}, \quad
b_{\alpha}^{(3)}  =  q^2 (q \varphi \varphi )_{\alpha} - 
q_{\alpha} \, (q \varphi \varphi q), $$
$\varphi_{\alpha \beta} =  
F_{\alpha \beta} \left / \sqrt{F_{\mu \nu}^2 / 2} \right .$ 
is the dimensionless tensor of the external 
magnetic field, 
${\tilde \varphi}_{\alpha \beta} = \frac{1}{2} \varepsilon_{\alpha \beta
\mu \nu} \varphi_{\mu \nu}$ is the dual tensor.
The indices of four-vectors and tensors inside the 
parentheses are contracted consecutively, e.g.
$( q \varphi )_{\alpha} = q_\beta \varphi_{\beta \alpha}$, etc.

It is well-known that only the photons, both real and virtual,
with the $\lambda = 2$ 
polarization participate in all the electron - photon processes in 
a strong magnetic field. Taking into account that the electron mass 
being the physical value is gauge independent, it is convenient to 
calculate it in the gauge where the photon propagator of the mode 2
takes the form
\begin{eqnarray}
{\cal D}^{(2)}_{\mu \nu} (k) = - i \; {\cal D} (k^2) \;
\tilde\varphi_{\mu \rho} \tilde\varphi_{\rho \nu}, \quad
{\cal D} (k^2) = \frac{1}{k^2 - {\cal P}^{(2)}} .
\label{eq:D_munu2}
\end{eqnarray}
To calculate the electron mass operator in the leading log approximation, 
it is enough to know the mode 2 photon polarization operator at the 
one-loop level~\cite{Shabad:1988}:
\begin{eqnarray}
{\cal P}^{(2)} = - \frac{2 \alpha}{\pi} \; e B \; 
H \left( \frac{k_{\mprl}^2}{4 m^2} \right) + 
\frac{\alpha}{3 \pi} \; k^2 \; \ln \frac{e B}{m^2}.  
\label{eq:P_2}
\end{eqnarray}
Here, the first term arises from virtual electrons at the ground Landau 
level, while the second term is determined by the higher Landau levels.
The function $H(k_{\mprl}^2/4 m^2)$ for $k_{\mprl}^2 < 0$ has the form
\begin{eqnarray}
H(z) = \frac{1}{2 \sqrt{z(z + 1)}} \ln 
\frac{\sqrt{z + 1} + \sqrt{z}}{\sqrt{z + 1} - \sqrt{z}} - 1, 
\quad z = \frac{|k_{\mprl}^2|}{4 m^2}.
\label{eq:H(z)}
\end{eqnarray}
For the region of parameters we are interested in, $|k_{\mprl}^2| \gg 4 m^2$, 
we have $H(z) \simeq -1$. Consequently, the first term in Eq.~(\ref{eq:P_2}) 
acquires the meaning of the effective photon mass squared, 
$m_\gamma^2 = (2 \alpha/\pi) e B$, induced by a magnetic field. 
Substituting~(\ref{eq:P_2}) into Eq.~(\ref{eq:D_munu2}), and multiplying
by $\alpha$, we obtain for the function $\alpha \, {\cal D} (k^2)$:
\begin{eqnarray}
\alpha \, {\cal D} (k^2) = 
\frac{\alpha}{k^2 - k^2 \; (\alpha/3 \pi) \; \ln (e B/m^2) 
- m_\gamma^2 (\alpha)} .
\label{eq:D_2}
\end{eqnarray}
It is seen from Eq.~(\ref{eq:D_2}) that the contribution of the higher 
Landau levels leads in fact to the electron charge renormalization 
in a strong magnetic field 
\begin{eqnarray}
\alpha \longrightarrow \alpha_R = 
\frac{\alpha}{1 - (\alpha/3 \pi) \; \ln (e B/m^2)}. 
\label{eq:alphaR}
\end{eqnarray}
Based on this, the Eq.~(\ref{eq:D_2}) can be presented in the form
\begin{eqnarray}
\alpha \, {\cal D} (k^2) = 
\frac{\alpha_R}{k^2 - m_\gamma^2 (\alpha_R)} .
\label{eq:D_2R}
\end{eqnarray}

By this means the mode 2 photon manifests itself in a strong magnetic field 
as a massive vector boson interacting with an electron with 
the renormalized coupling constant $\alpha_R$. 

For the $n$-loop contribution into the electron mass operator in the 
leading log approximation, which is defined by the rainbow diagram,
we have obtained
\begin{eqnarray}
\Delta M^{(n)} = m \, \left[\frac{\alpha_R}{2 \pi} \, 
\left(\ln \frac{\pi}{\alpha_R} - \gamma_E \right) \,
\ln \frac{e B}{m^2} \right]^n. 
\label{eq:DM_n}
\end{eqnarray}

Finally, for the electron mass operator in a strong magnetic field 
one obtains
\begin{eqnarray}
M = m \left/ 
\left[1 - \frac{\alpha_R}{2 \pi} \, 
\left(\ln \frac{\pi}{\alpha_R} - \gamma_E \right) \,
\ln \frac{e B}{m^2} \right]
\right ..
\label{eq:M_our}
\end{eqnarray}
It should be noted that our expression~(\ref{eq:M_our}) 
for the electron mass operator in the single logarithm field asymptotics 
is different from the analogous result obtained 
in Ref.~\cite{Loskutov:1983}, which had the form
\begin{eqnarray}
M = m \; \exp \left(\frac{\alpha}{2 \pi} \, 
\ln \frac{\pi}{\alpha} \, \ln \frac{e B}{m^2} \right).
\label{eq:M_Skob}
\end{eqnarray}

To the first order in the parameter $\alpha \ln (e B/m^2)$, 
the results~(\ref{eq:M_our}) and~(\ref{eq:M_Skob}) almost 
coinside~\footnote{The Euler constant $\gamma_E$ is lost in 
Ref.~\cite{Loskutov:1983}.}
and are reduced to our formula~(\ref{eq:DM_1}). 
However, already in the second order in $\alpha \ln (e B/m^2)$ 
the results are different because of two reasons.
First, the arithmetic error was made in Ref.~\cite{Loskutov:1983} 
in the calculation of the $n$-loop rainbow diagram which led 
to an incorrect summation of the perturbation expansion. 
Second, the authors~\cite{Loskutov:1983} did not take into account 
the contribution of the higher Landau levels which is essential 
in the logarithm asymptotics. The fundamental difference of our 
result~(\ref{eq:M_our}) from the result~(\ref{eq:M_Skob}) of 
Ref.~\cite{Loskutov:1983} is in the fact that the electron mass 
as the function of the field strength, $M(B)$, is the singular one, 
because the denominator of Eq.~(\ref{eq:M_our}) can, in principle, 
tend to zero. It should be noted, however, that the singular 
behaviour of the function $M(B)$ could manifest itself only in 
a superstrong field $B \sim 10^{75}\;$ G.  

\bigskip

{\bf Acknowledgements}  

We are grateful to V.A. Rubakov and to the participants 
of the seminar of the Theoretical Department 
of the Institute for Nuclear Research of RAS 
for fruitful discussion.

This work was supported in part by the Russian Foundation for Basic 
Research under the Grant N~01-02-17334
and by the Ministry of Education of Russian Federation under the 
Grant No. E00-11.0-5.

\vspace{30mm}

%---------------------------------------------------------
%---------------------------------------------------------


\newpage

\begin{thebibliography}{99}
%
\bibitem{Jancovici:1969} 
   B. Jancovici,
   {\it Phys. Rev.} {\bf 187}, 2275 (1969).
%
\bibitem{Loskutov:1976} 
   Yu. M. Loskutov, V. V. Skobelev, 
   {\it Phys. Lett.} {\bf A 56}, 151 (1976). 
%
\bibitem{Loskutov:1979} 
   Yu. M. Loskutov, V. V. Skobelev, 
   {\it Teor. Mat. Fiz.} {\bf 38}, 195 (1979). 
%
\bibitem{Loskutov:1981} 
   Yu. M. Loskutov, V. V. Skobelev, 
   {\it Teor. Mat. Fiz.} {\bf 48}, 44 (1981). 
%
\bibitem{Gusynin:1999} 
   V. P. Gusynin, A. V. Smilga,
   {\it Phys. Lett.} {\bf B 450}, 267 (1999). 
%
\bibitem{Loskutov:1983} 
   Yu. M. Loskutov, V. V. Skobelev, 
   {\it Vestn. Mosk. Univ., Fiz., Astron.} {\bf 24}, 95 (1983). 
%
\bibitem{Shabad:1988}
   A. E. Shabad, 
   {\it Tr. Fiz. Inst. Akad. Nauk SSSR} {\bf 192}, 5 (1988).
%
\end{thebibliography}
\end{document}